\documentclass[letterpaper]{article} 
\usepackage{aaai24}  
\usepackage{times}  
\usepackage{helvet}  
\usepackage{courier}  
\usepackage[hyphens]{url}  
\usepackage{graphicx} 
\usepackage{natbib}  
\usepackage{caption} 
\usepackage{algorithm}
\usepackage{algorithmic}

\usepackage{amsthm,amsmath,amssymb}
\usepackage{amsfonts}       
\usepackage{booktabs}       
\usepackage[T1]{fontenc}    
\usepackage{graphicx} 
\usepackage[utf8]{inputenc} 
\usepackage{microtype}      
\usepackage{multirow}

\usepackage{nicefrac}       
\usepackage{pgfplots}
\usepackage{pifont}
\usepackage{url}            
\usepackage{tikz}
\usepackage{xcolor}         

\usepackage{newfloat}
\usepackage{listings}
\DeclareCaptionStyle{ruled}{labelfont=normalfont,labelsep=colon,strut=off} 
\floatstyle{ruled}
\newfloat{listing}{tb}{lst}{}
\floatname{listing}{Listing}
\newtheorem{definition}{Definition}
\newtheorem{example}{Example}

\newtheorem{theorem}{Theorem}

\newcommand{\new}[1]{\textcolor{black}{#1}}

\newcommand{\cmark}
{{\ding{52}}}
\newcommand{\xmark}{{\ding{53}}}

\title{Stable Matchings in Practice: A Constraint Programming Approach}
\author{
    Zhaohong Sun\textsuperscript{\rm 1, 2},
    Naoyuki Yamada\textsuperscript{\rm 2},
    Yoshihiro Takenami\textsuperscript{\rm 2},
    Daisuke Moriwaki\textsuperscript{\rm 2},
    Makoto Yokoo\textsuperscript{\rm 1}
}
\affiliations{
    \textsuperscript{\rm 1} Kyushu University, Japan\\
    \textsuperscript{\rm 2} CyberAgent Inc, Japan
}

\usepackage{bibentry}

\begin{document}

\maketitle

\begin{abstract}
  We study a practical two-sided matching problem of allocating children to daycare centers, which has significant social implications. We are cooperating with several municipalities in Japan and our goal is to devise a reliable and trustworthy clearing algorithm to deal with the problem. In this paper, we describe the design of our new algorithm that minimizes the number of unmatched children while ensuring stability. We evaluate our algorithm using real-life data sets, and experimental results demonstrate that our algorithm surpasses the commercial software that currently dominates the market in terms of both the number of matched children and the number of blocking coalitions (measuring stability). Our findings have been reported to local governments, and some are considering adopting our proposed algorithm in the near future, instead of the existing solution. Moreover, our model and algorithm have broader applicability to other important matching markets, such as hospital-doctor matching with couples and school choice with siblings.
\end{abstract}

\section{Introduction}

 With the prevalence of dual-career households in recent years, the demand for daycare facilities in Japan, especially in metropolitan areas, has soared. Unfortunately, scarce space and insufficient teachers lead to a long waiting list each year, leaving numerous children unable to enroll in daycare centers. The waiting child problem becomes one of the critical social challenges nowadays (see the press conference by the Japanese Prime Minister on March 17, 2023\footnote{\url{https://japan.kantei.go.jp/101_kishida/statement/202303/_00015.html}}).

The allocation of children to daycare centers in Japan is not done on a first-come, first-served basis. Instead, the allocation process starts with families submitting applications to their local government office, which contains basic information such as children's age, guardians' health conditions and work schedule, as well as preferences over acceptable daycare centers. Each municipality adopts a unique scoring system that strictly prioritizes children. The scoring system is designed in a way that children who may have greater needs or face additional challenges have more chances of utilizing daycare services. Typically, children from low-income or single-parent households, and those whose guardians are suffering from diseases or disabilities take precedence over others. Some local governments formulate lottery rules for tie-breaking when children have identical priority scores. The allocation is then computed by a centralized matching algorithm that allows for both families' preferences over daycares and daycares’ priorities over children. 

The daycare matching process has faced criticism due to long waiting lists and the Japanese government has made considerable efforts to address this issue, including improving working conditions and increasing salaries for childcare workers to engage more people in early childhood education, and providing financial assistance for families in need to enlarge their options of affordable daycare centers.    
Although the number of children on the waiting list significantly decreased recently, the shortage of daycare facilities continues. This is because not all unmatched children are counted in the waiting list, such as those who live near daycare centers with vacant slots but are only willing to attend certain oversubscribed daycare centers and those whose parents have to suspend their jobs or extend their childcare leave.
Thus, despite these measures, the waiting child problem remains a major social challenge and long waiting lists have a profound impact on young couples’ careers and lives.

\begin{table*}[tb]
\centering
\resizebox{1.6\columnwidth}{!}{
\begin{tabular}{ccccccc}
 \hline
   & \normalsize{initial} & \multicolumn{3}{c}{\normalsize{\# siblings}} & \normalsize{any joint} & \normalsize{transferable}  
   \\
   & enrollments & $2$ & $3$ & any & preferences & quotas
    \\
     \hline
 \citep{McMa10a} & \xmark & \cmark & & & \xmark & \xmark
 \\
  \hline
 \citep{ABH14a} & \xmark & \cmark & & & \cmark & \xmark
 \\
 \hline
  \citep{MMT17a} & \xmark & \cmark & & & \cmark  & \xmark
 \\
   \hline
 \citep{STY+23a} & \cmark & \xmark & & & \xmark & \xmark
 \\
 \hline
 \citep{Okum19a} & \xmark & \xmark & & & \xmark & \cmark
 \\
 \hline  
 \citep{KaKo23a} & \xmark & \xmark & & & \xmark & \cmark
 \\
 \hline  
  \citep{DMP22a} & \xmark & \cmark & & & \xmark & \xmark  
 \\
 \hline  
  \citep{CEE+22a} & \cmark &  & & \cmark & \xmark  & \xmark
 \\
 \hline  
  \citep{STM+23a} & \cmark & & \cmark &  & \cmark & \xmark
 \\
 \hline  
   This Work & \cmark & &  & \cmark & \cmark & \cmark
 \\
 \hline  
\end{tabular}
}
\caption{Comparison with some recent papers. \cmark indicates that the proposed approach in that paper is applicable to a particular feature.}
\label{table:comparison}
\end{table*}

The daycare matching problem bears similarities with conventional two-sided matching problems such as college admissions and job hunting, where one side of the market consists of families who submit applications on behalf of their children and the other side consists of daycare centers with limited resources (e.g., room space, teachers). However, the daycare matching market possesses three features that set it apart from classical matching models. These features include i) \emph{transfers} (i.e., some children who are already enrolled prefer to be transferred to other daycare centers), ii) \emph{siblings} (i.e., several children from the same family report joint preferences and only consent to an assignment if all of them are matched),  and iii) \emph{transferable quotas} (i.e., a daycare center may partition grades into grade groups and available spots can be used by any child within the same grade group).
It is well-known that when couples exist, there may not exist any matching satisfying stability \citep{Roth84a} (i.e., one of the most important solution concepts in matching theory \citep{Roth08a}) and determining whether there exists a stable matching is NP-complete \citep{Ronn90a}. The presence of these complexities poses more significant challenges. 

The objective of this research is to develop a trustworthy algorithm to help municipalities tackle the waiting children problem. The key research question is \emph{how to design and implement practical matching algorithms that minimize the number of children on the waiting list in a transparent, stable and computationally efficient manner.}

We next explain why these principles need to be taken into consideration. 
\emph{Transparency} requires that the public and policymakers can understand and verify how algorithms make decisions. Currently, the daycare matching system that dominates the market does not disclose its code for commercial purposes. Thus, it is unclear why a certain child is assigned to a particular daycare center or why some children are unmatched. 
\emph{Stability}  plays a critical role in the success of many real-world applications, 
including hospital-residency matching programs and public school choice. In the context of daycare matching, stability can be decomposed into fairness and non-wastefulness, both of which are vital requirements mandated by municipalities. 
\emph{Computational efficiency} of the matching algorithm is also important, as the number of children participating in the market is large. It usually costs government officers several weeks to calculate and verify the outcomes manually, the process of which is also prone to mistakes. Implementing the algorithm on a computer can yield an outcome in a few seconds or minutes, which would be much more efficient.

Our contributions are summarized as follows: Firstly, we formalize three particular features of the market and develop a comprehensive model that encompasses other important matching markets. 
\new{
Secondly, we present some new computational complexity results, including that it is NP-complete to check the existence of feasible and individual rational matching that differs from the initial enrollments. 
Thirdly, we propose a practical algorithm based on constraint programming (CP) which is a powerful technique for coping with NP-hard problems. Fourthly, we evaluate the effectiveness of our algorithm by conducting experiments on real-world data sets and summarize interesting findings on the factors that could increase the number of matched children. Lastly, we release our implementation that could benefit future work in related domains.
}

\section{Related Work}
There is a large body of literature on matching problems under preferences and we give a more detailed review of related work in Appendix. We next compare our work with some recent papers that also consider some of the features mentioned above. 
The daycare matching problem can be seen as a generalization of hospital-doctor matching with couples, where pairs of doctors participate in the market and aim to secure a pair of positions \citep{KPR13a,BMM14a,NgVo18a}. 
While some papers focus on school choice with initial enrollment, they do not take siblings into consideration \citep{HHKS+17a,STY18a,STY+23a}.
On the other hand, two recent papers delve into school choice with siblings, but assuming restrictive preferences of  families \citep{DMP22a, CEE+22a}.
Regarding the Japanese daycare matching problem, two prior works investigate transferable quotas but do not consider initial enrollments or siblings in their models ~\citep{Okum19a,KaKo23a}. Another study also examines the Japanese daycare matching market; however, they do not allow for transferable quotas and propose an algorithm suitable for a specific scenario with a maximum of three children per family \citep{STM+23a}.
Several papers tackle the practical daycare matching problem in European countries, but it is important to note that the settings and objectives of these studies differ significantly from ours \citep{VBPL17a,GJR+21a,RKG21a}.

\section{Model}
In this section, we present our model in the context of daycare matching. It can also be used to describe other matching markets such as hospital-doctor matching with couples, school choice with siblings or a general setting where multiple agents form a group and the group has a joint preference over the assignments of group members. 

An instance $I$ of the daycare matching problem consists of a tuple $I$ $=$ $(C$, $F$, $D$, $\omega$, $\succ_F$, $\succ_C$, $\succ_D$, $G$, $Q$). 
The set of children $C$ is partitioned into a set of disjoint families $F$, where each child $c$ belongs to exactly one family $f(c) \in F$. We denote the set of children associated with family $f$ as $C(f)$ and we assume children $C(f) = \{c_1, c_2, \cdots, c_k\}$ are \emph{sorted in a fixed order}, such as by age. 
The set of daycares is denoted as $D$, which includes a dummy daycare $d_0$ that represents the option of being unmatched for children. 
Some children may be  initially enrolled at a daycare $d\in D\setminus \{d_0\}$ and prefer to be transferred to a different one. We use the function $\omega(c) \in D$ to denote the initial daycare of child $c$. If $\omega(c) = d_0$, then child $c$ is considered a new applicant. We extend the concept of $\omega(c)$ to $\omega(f)$ for a family $f$ with children $C(f) = \{c_1, \ldots, c_k\}$ by defining $\omega(f) = (\omega(c_1), \ldots, \omega(c_k))$ as the initial enrollments of family $f$.

Each family $f$ has a strict preference ordering $\succ_f$ over tuples of daycare centers, representing the joint preferences of children $C(f)$. For example, if family $f$ has $k$ children $C(f)$ $=$ $\{c_1$, $c_2$, $\cdots$, $c_k\}$, then a tuple of daycares in $\succ_f$, denoted as $(d_1^*, d_2^*, \cdots, d_k^*)$, 
indicates that child $c_i$ attends daycare $d_i^*$ for each $i\in [1, k]$. It is possible for $d_1^* = d_2^*$, indicating that children $c_1$ and $c_2$ prefer to attend the same daycare.
If at least one child $c \in C(f)$ is initially enrolled at a daycare $d \in D\setminus \{d_0\}$, then the initial enrollments $\omega(f)$ are appended to the end of $\succ_f$.
%
For child $c$ from family $f$, the notation $\succ_c$ denotes a projected preference ordering of child $c$ over daycares that is generated from $\succ_f$. If family $f$ has $k$ children $C(f)=\{c_1, \cdots, c_k\}$, then the projected preference ordering $\succ_{c_i}$ of child $c_i$ consists of the $i$-th element in each tuple of daycares in $\succ_f$, as shown in Example~\ref{example:projected_preferences}. It is important to note that $\succ_c$ does not represent child $c$'s individual preferences, and the same daycare may occur multiple times in $\succ_{c}$.

Each daycare $d$ has a strict priority ordering over children denoted by $\succ_d$. We assume children who are initially enrolled at daycare $d$ have higher priority by $\succ_d$ than others who are not. This is a realistic assumption and it is confirmed by multiple municipalities.

\begin{example}[Instance of Projected Preferences]
\label{example:projected_preferences}
Consider three daycares $D = \{d_0, d_1, d_2\}$ and 
one family $f$ with two children $C(f)=\{c_1, c_2\}$. Family preferences $\succ_f$ and projected preferences of children $c_1$ and $c_2$ are as follows:
\vspace{-0.2mm}
\begin{align*}
& \succ_f: (\textcolor{black}{d_1}, \textcolor{black}{d_1}), (\textcolor{black}{d_2}, \textcolor{black}{d_2}), (\textcolor{black}{d_1}, \textcolor{black}{d_0}), (\textcolor{black}{d_2}, \textcolor{black}{d_0}), (\textcolor{black}{d_0}, \textcolor{black}{d_1}), (\textcolor{black}{d_0}, \textcolor{black}{d_2})
\\
& \succ_{c_1}: d_1, d_2, d_1, d_2, d_0, d_0
\\
& \succ_{c_2}: d_1, d_2, d_0, d_0, d_1, d_2.
\end{align*}
\end{example}

There is a set of grades $G$ and each child $c$ is associated with one grade denoted by $G(c)$. Let $Q$ represent grade-specific quotas at all daycares, and specifically $Q(d, g)$ represents the quota for grade $g$ at daycare $d$. 
A daycare center may partition grades $G$ into disjoint grade groups and available spots can be used by any child within the same grade group, which allows for greater flexibility in the matching process.  
Formally, for each daycare $d$, the set 
$G(d) = \{G_{d_1}, \cdots, G_{d_t}\}$
is a partition of grades $G$ into disjoint grade groups.
%
Given a grade group $G_d \in G(d)$ at daycare $d$, let $Q(d, G_d) =\sum_{g\in G_d} Q(d, g)$ denote a transferable quota, which is the sum of grade-specific quotas of each grade $g\in G_d$.
The use of transferable quotas has been proposed in previous work and it is employed by some daycare centers in practice. 
Take Tama city for example. The majority of slots are assigned at the start of April annually, comforting to rigid grade-specific quotas. Each month, there are only a limited number of available slots, and a few daycares allow transferable quotas. 
\begin{example}[Transferable Quotas]
    \label{example:transferable_quotas}
    Suppose daycare $d$ partitions grades $G$ into $G(d) = \{ \{0\}, \{1,2\}, \{3,4,5\}\}$. It indicates that grade $0$, grades $\{1, 2\}$ and grades $\{3, 4, 5\}$ form three grade groups respectively, where grade-specific quotas are transferable within each grade group.  
If daycare $d$ does not permit transferable quotas, then each grade itself is considered a grade group, i.e., $G(d) = \{ \{0\}, \{1\}, \{2\}, \{3\},\{4\},\{5\}\}$. 
\end{example}
An outcome, denoted by $\mu$, represents a matching between children $C$ and daycares $D$ such that each child $c$ is matched to one daycare $d\in D$. We use $\mu(c)$ to denote the daycare matched to child $c$ in the outcome $\mu$. Similarly, $\mu(d)$ represents the set of children matched to daycare $d$ in $\mu$. To specify the set of children of a particular grade $g$ who are matched to daycare $d$ in $\mu$, we use the notation $\mu(d,g)$. 
Furthermore, $\mu(d, G_d)$ refers to the set of children belonging to the grade group $G_d$ who are matched to daycare $d$ in $\mu$. Additionally, $\mu(f)$ is a tuple of daycares matched to children $C(f)$ in $\mu$. Note that $C(f) = \{c_1,\ldots,c_k\}$ is sorted in a predetermined order, and $\mu(f) = (d_1^*,\ldots,d_k^*)$ indicates that child $c_i$ is matched with daycare $d_i^*$, respectively.

\section{Properties}

In this section, we present several desirable properties that an algorithm should satisfy. These properties have not only been extensively studied in matching theory, but are also practical requirements in real-life matching markets. 
The first fundamental property is individual rationality, which states that each family must be matched to some assignment that is weakly better than its initial enrollments. 
\begin{definition}[Individual Rationality]
\label{def:IR}
An outcome $\mu$ satisfies individual rationality if for each family $f \in F$, either $\mu(f) \succ_f \omega(f)$ or $\mu(f) = \omega(f)$ holds.
\end{definition}




The second fundamental property, feasibility in Definition~\ref{def:feasibility}, has two requirements. The first condition is that each child must be matched to exactly one daycare, as any remaining unmatched children can be assigned to the dummy daycare $d_0$. The second condition is that for each daycare $d$, the number of matched children from grade group 
$G_d$ cannot exceed the transferable quota $Q(d, G_d)$ for that grade group.
This ensures that the matching complies with the capacity of each daycare.

\begin{definition}[Feasibility]
\label{def:feasibility}
An outcome $\mu$ is feasible (w.r.t. transferable quotas) if 
\\
i) $\forall c\in C, |\mu(c)| = 1$; and 
\\
ii) 
$\forall d \in D, \forall \ G_d \in G(d), |\mu(d, G_d)| \leq Q(d, G_d)$.
\end{definition}

At its core, stability refers to a state in which no two agents could both benefit from leaving their current match and forming a new one \citep{GaSh62a}.
To apply this idea to our setting, we next introduce a choice function of daycares, which simplifies the representation of stability. 

\begin{definition}[Choice Function $Ch_d$]
    \label{def:choice_func}
    Given a set of children $C'\subseteq C$, the choice function $Ch_d$ of daycare $d$ is defined as follows:  
    For each grade group 
    $G_d \in G(d)$, let $C'({G_d})$ 
    denote the set of children in $C'$ that belong to that grade group. The choice function then selects children from $C'$ one at a time, based on daycare priority ordering $\succ_d$, without exceeding the transferable quota 
    $Q(d, G_d)$ for each grade group $G_d$.
\end{definition}

\begin{example}[Choice Function]
\label{example:choice}
Consider four children denoted as $C=\{c_1, c_2, c_3, c_4\}$ with respective ages of $0, 0, 3, 4$.
Suppose there exists one
daycare $d$ which partitions grades $G$ into $G(d) = \{ \{0\}, \{1,2\}, \{3,4,5\}\}$.
The specific quotas for each age from 0 to 5 are as follows: $[1, 1, 1, 2, 0, 1]$.  The priority ordering of the children is given by $\succ_d: c_1, c_2, c_3, c_4$.

Based on the choice function defined in Definition~\ref{def:choice_func}, the selected children are $c_1$, $c_3$, and $c_4$. This is due to the fact that there is a transferable quota of $1$ for age $0$, as well as a transferable quota of $3$ for the age group $\{3,4,5\}$.
\end{example}

We generalize stability in
Definition~\ref{def:stability}, which is characterized as the absence of blocking coalitions. Briefly speaking, a matching $\mu$ is stable if there is no family $f$ and a tuple of daycares $\succ_{f, j}$ (the $j$-th tuple in $\succ_f$) that could form a blocking coalition. A blocking coalition occurs when $f$ prefers $\succ_{f, j}$ over its current match in $\mu$, and the daycares involved in $\succ_{f, j}$ can accommodate all children in $C(f)$, along with a set of children who have higher priority than at least one child in $C(f)$, without violating the feasibility constraints.
Definition~\ref{def:stability} captures the essentials of stability and coincides with some previous concepts for the setting of matching with couples (e.g., \citep{ABH14a}). 

Definition~\ref{def:stability} is also similar to the stability for the matching with contracts model \citep{HaMi05a} where no additional contract is preferred by both families and daycares.

\begin{definition}[Stability]
\label{def:stability}
Given a feasible outcome $\mu$, family $f$ with children 
$C(f) = \{c_1, \cdots, c_k\}$ and a tuple of daycares $\succ_{f, j} = (d^*_1, \ldots, d^*_k)$ located at position $j$ of $\succ_f$,  will form a blocking coalition if i) family $f$ prefers 
$\succ_{f, j}$ over $\mu(f)$, 
and 
ii) $\forall d\in D(f,j)$, $C(f, j, d) \subseteq Ch_d(\mu(d) \cup C(f, j, d))$ where $D(f, j)$  denotes the set of distinct daycares located at position $j$ of $\succ_f$ and $C(f, j, d)$ denotes a subset of children who apply to daycare $d$ w.r.t. $\succ_{f, j}$.
%
A feasible matching is stable if there is no blocking coalition.
\end{definition}

\begin{example}[Stability]
\label{example:stability}
Consider two families $f_1$ with children $C(f_1) = \{c_1, c_2\}$ and $f_2$ with Children $C(f_2) = \{c_3\}$. There are two daycares $D=\{d_1, d_2\}$ with one slot each. All children have the same age. 
The preferences and priorities are as follows: 
\begin{align*}
    & \succ_{f_1}: (d_1, d_2) \quad \succ_{d_1}: c_1, c_3
    \\
    & \succ_{f_2}: d_1, d_2 \qquad \succ_{d_2}: c_3, c_2.
\end{align*}
Consider matching $\mu_1$ where 
family $f_1$ is matched to $(d_1, d_2)$ while $f_2$ is unmatched, i.e., $\mu_1(c_1) = d_1$, $\mu_1(c_2) = d_2$, $\mu_1(c_3) = \emptyset$. 
Matching $\mu_1$ is not stable, because family $f_2$ can form a blocking coalition with daycare $d_2$. 

More specifically, i) family $f_2$
prefers $d_2$ over $\emptyset$ and ii) the notion $C(f_2, 2, d_2) = \{c_3\}$ (i.e., the set of children from family $f_2$ who applies to daycare $d_2$ w.r.t. the second element in $\succ_{f_2}$), and $\{c_3\} = Ch_{d_2}(\mu_1(d_2) \cup \{c_3\})$. 
\end{example}

Depending on whether certain children are excluded from $\bigcup_{d\in D(f,j)} \mu(d)$, blocking coalitions in Definition~\ref{def:stability} can be classified into two categories: justified envy (when some children are removed to create availability) or waste (when there is still enough capacity). Formally,


\begin{definition}[Decomposition of Blocking Coalition]
\label{def:block}
Given a feasible outcome $\mu$, family $f$ with children $C(f) = \{c_1, \cdots, c_k\}$ and a tuple of daycares $\succ_{f, j} = (d^*_1, \ldots, d^*_k)$ located at position $j$ of $\succ_f$ form a blocking coalition.

Let $Re(f, j) = \bigcup_{d\in D(f, j)} \mu(d) \setminus Ch_d(\mu(d) \cup C(f, j, d))$ denote a set of children who are matched to $d\in D(f, j)$ but are not chosen by $Ch_d(\mu(d) \cup C(f, j, d))$, i.e., a set of children who are removed  to make room for children $C(f)$. 
\begin{itemize}
    \item If $Re(f, j) \neq \emptyset$, then family $f$ has justified envy toward $Re(f, j)$.
    \item If $Re(f, j) = \emptyset$, then family $f$ claims that outcome $\mu$ is wasteful.
\end{itemize}
A feasible matching is fair if it is free of justified envy. A feasible matching is non-wasteful if no family $f$ claims the matching is wasteful.
\end{definition}

Fairness is mandated in the daycare matching process under Japanese laws and regulations. 
In the context of daycare matching (or more generally in two-sided matching markets), fairness is commonly defined as elimination of justified envy. That is, one child with a higher priority score has justified envy toward another child with a lower priority score if the latter is matched to a more preferred daycare center. This fairness concept is widely employed across the country.

Non-wastefulness is advocated in the government plan, suggesting that in addressing the waiting child problem, it's important to make full use of all available resources, which could help to alleviate the shortage of daycare spaces. 
It states that no family cannot be matched to a more preferred tuple of daycare centers without changing the assignment of any other family. 

 Both concepts coincide with their counterparts in the classical school choice without siblings and initial enrollments (e.g., ~\cite{AbSo03b}).

\begin{example}[Decomposition of Blocking Coalition]
\label{example:decomposition}
Consider the instance in Example~\ref{example:stability}.
Matching $\mu_1$ does not satisfy fairness, because family $f_2$ has justified envy toward child $c_2$.
Consider matching $\mu_2$ where 
family $f_1$ is unmatched while $f_2$ is matched to $d_2$, i.e., $\mu_1(c_1) = \emptyset$, $\mu_1(c_2) = \emptyset$, $\mu_1(c_3) = d_2$. 
Matching $\mu_2$ does not satisfy non-wastefulness, because family $f_2$ be matched to daycare $d_1$ without affecting any other family. More specifically, $f_2$ could form a blocking coalition with daycare $d_1$ and $Re(f_2, 1) = \emptyset$.
\end{example}

An alternative non-wastefulness concept was introduced \citep{STM+23a}, allowing siblings to exchange their daycare assignments at the expense of daycares' welfare. We implemented a different stability concept that incorporated the revised non-wastefulness concept in our experiments. However, 
it does not affect the stable matching across all the datasets we examined.

\section{Computational Complexity}

This section is devoted to the computational complexity of verifying whether outcomes with certain desirable properties exist. Previous studies have shown that identifying stable outcomes in matching problems involving couples is NP-hard, and this complexity result also applies to the daycare matching problem. However, the inclusion of initial enrollments makes the problem more challenging. We demonstrate that determining the existence of a feasible and individually rational outcome that is different from the initial matching is NP-complete, even under restrictive conditions.
\begin{theorem}
\label{theo:IR:NP-C}
It is NP-complete to check whether there exists a feasible and individually rational outcome that differs from the initial matching, even if i) each family has at most two children; ii) the length of each family's preference ordering is at most $2$.
\end{theorem}
\begin{proof}
The problem of finding a matching that is different from the initial matching and satisfies feasibility and individual rationality is in the NP complexity class,
as we can guess any matching and verify whether it satisfies these requirements in polynomial time. 

To further demonstrate the problem is NP-hard, we will show a polynomial-time reduction from the subset sum problem, which is known to be NP-complete  \citep{GaJo79a}. In the subset sum problem, we are given a set of integers $V=\{v_1, \cdots, v_n\}$ and a target $\alpha$, and the question is to determine whether there exists a subset $V'\subseteq V$ such that the sum of its elements equals $\alpha$, i.e., $\sum_{v\in V'} v = \alpha$. 
Let $\beta = \sum_{v\in V} v$ denote the sum of all integers from $V$. 

We can construct a corresponding instance of daycare matching and show that there exists a feasible and individually rational outcome that differs from the initial matching if and only the subset sum problem admits a yes-instance. Due to space limitation, more details are presented in Appendix.
\end{proof}
\vspace{-3mm}
Pareto optimality is another useful concept in economics that is stronger than non-wastefulness and it is incompatible with stability in general \citep{RoSo92a}. We say an outcome $\mu$ is Pareto optimal if there does not exist any other matching $\mu'$ s.t. all families weakly prefer $\mu'$ over $\mu$ and at least one family strictly prefers $\mu'$ over $\mu$.
 Although it is not the focus of this paper, the following theorem shows that it is co-NP-complete to verify whether a feasible matching satisfies Pareto optimality. 
\begin{theorem}
\label{theorem:PO}
    It is co-NP-complete to check whether a feasible and individually rational matching satisfies Pareto optimality, even if i) each family has at most two children; ii) the length of each family's preference ordering is at most $2$.
\end{theorem}

\begin{proof}
Consider the induced instance of daycare matching in the proof of Theorem~\ref{theo:IR:NP-C}. We next show that checking whether the initial matching satisfies Pareto optimality is co-NP-complete. 
The complement problem is to check whether there exists a feasible matching that Pareto dominates the initial one. Note that any feasible and individually rational matching other than the initial one Pareto dominates the initial matching. From Theorem~\ref{theo:IR:NP-C}, we know this problem is NP-complete and thus the complement problem is co-NP-complete. This completes the proof of Theorem~\ref{theorem:PO}.
\end{proof}

\section{Algorithm Design}

In this section, we present our new algorithm that aims to address the daycare matching problem with three particular features. Given the computational intractability results in the preceding section, we utilize constraint programming (CP), a powerful technique for solving NP-hard problems, to develop an efficient and practical solution. 
Another important reason that we choose the CP solution is to reduce the cost of social implementation. We can represent the problem as a mixed-integer linear program (MILP) and invoke successful solvers for MILPs, which are commercial and expensive. On the other hand, Google OR-Tools is a free and open-source software suite and provides a  powerful CP-SAT solver.

As a stable outcome may not always be theoretically guaranteed, our CP algorithm is specifically designed to prioritize two main objectives. First, it aims to identify the minimum number of blocking coalitions, and subsequently, find a matching that maximize the number of matched children with the minimum number of blocking coalitions. This idea is inspired by \citep{MMT17a}.

We next introduce a set of variables that are used in the CP model. These variables are carefully defined and utilized to facilitate the optimization process.

For each family $f\in F$ and each position $p\in [1, \mid\succ_f\mid]$, let $x[f, p]$ denote a binary variable indicating whether family $f$ is matched to the $p$-th assignment in preference ordering $\succ_f$. 
%
\begin{equation}
\label{var:x(f, p)}
x[f, p] = 
\begin{cases}
1 \quad \small{\text{if $f$ is matched to the $p$-th assignment in} \succ_f}
\\
0 \quad \small{\text{otherwise.}}
\end{cases}
\end{equation}
For each child $c$ from family $f$ and each position $p$, let $x[c, p]$ denote a binary variable that equals $x[f, p]$, indicating whether child $c$ is matched to the $p$-th element in child $c$'s projected preference $\succ_c$. Note that we use $x[c, p]$ for illustration purposes only, and 
do not necessarily create these variables in our algorithm. 
\begin{equation}
\label{cons:x(c, p)}
x[c, p] = x[f, p]
\end{equation}
For each child $c$ from family $f$ and each daycare $d$ that appears in projected preference ordering $\succ_c$, create an integer variable $x[c, d]$ indicating whether child $c$ is matched to daycare $d$, where $P(c,d)$ denotes a set of positions corresponding to daycare $d$ in $\succ_{c}$. Using variable $x[c, d]$ is more convenient for calculating the number of children matched to a particular daycare $d$.
\begin{equation}
\label{var:x:c}
x[c, d] = \sum\nolimits_{p \in P(c, d)} x[c, p]
\end{equation}
For each family $f$ and each position $p$ in $\succ_f$, create an integer variable $\alpha[f, p]$. This variable indicates whether family $f$ is matched to the $i$-th assignment in $\succ_f$, where $i$ is no larger than $p$. In other words, it indicates whether family $f$ is matched to a weakly better assignment than the $p$-th assignment.
\begin{equation}
\label{var:b}
\alpha[f, p] = \sum\nolimits_{i = 1}^p x[f, i]
\end{equation}
Before introducing more variables, we need to explain several new notations. Recall that $\succ_{f, p}$ denotes the $p$-th tuple of daycares in $\succ_f$ and $D(f, p)$ denotes the set of distinct daycares appearing in $\succ_{f, p}$. 
Let $C(f, p, d, G_d)$ 
denote a set of children who meet the following criteria: i) the child is from family $f$, ii) the child applies to daycare $d$ with respect to $\succ_{f,p}$, and iii) the child has some grade $g$ that falls into the grade group $G_d$ at daycare $d$, i.e., $g \in G_d$.
Let 
$C(f, p, d) = \bigcup_{G_d \in G(d)} C(f, p, d, G_d)$
denote a set of children from family $f$ who apply to daycare $d$ w.r.t. $\succ_{f,p}$. 
Additionally, 
let $C^{+}(f, p, d, G_d)$ denote a set of children who i) are \emph{not} from family $f$, ii) have higher priority than at least one child in $C(f, p, d, G_d)$ by priority ordering $\succ_d$, and iii) have some grade $g \in G_d$.
\\
We create a binary variable 
$\gamma[f, p, d, G_d]$ for each family $f$, each position $p$ in $\succ_f$, each daycare $d \in D(f, p)$, and each grade group $G_d \in G(d)$. This variable indicates whether daycare $d$ can accommodate children $C(f, p, d, G_d)$ along with $C^{+}(f, p, d, G_d)$ without exceeding transferable quota $Q(d, G_d)$ for grade group $G_d$ at daycare $d$. 
\begin{equation}
\label{var:gamma:d:g}
\gamma[f, p, d, G_d] = 
\begin{cases}
1 \quad \text{if } 
\sum_{c\in C^{+}(f, p, d, G_d)} 
x[c, d] \\ \qquad + |C(f, p, d, G_d)| \leq Q(d, G_d)
\\
0 \quad\text{otherwise.}
\end{cases}
\end{equation}
For each family $f$, each position $p$ in $\succ_f$, and each daycare $d \in D(f, p)$, create a binary variable $\gamma[f, p, d]$ indicating whether daycare $d$ can accommodate children $C(f, p, d)$ along with children $\bigcup_{G_d\in G(d)} C^{+}(f, p, d, G_d)$ simultaneously without exceeding transferable quotas for all grade groups.
\begin{equation}
\label{var:gamma:d}
\gamma[f, p, d] = 
\begin{cases}
1 \quad 
\text{if } \ \forall G_d\in G(d), \  \gamma[f,p,d,G_d] = 1
\\
0 \quad\text{otherwise.}
\end{cases}
\end{equation}
For each family $f$ and each position $p$ in $\succ_f$, create a binary variable $\gamma[f, p]$ indicating whether daycares $D(f, p)$ will choose children $C(f)$ simultaneously. That is, $\gamma[f, p]$ equals $1$ if and only if $\gamma[f,p,d]$ equals $1$ for each distinct $d \in D(f, p)$.
\begin{equation}
\label{var:gamma}
\gamma[f, p] = 
\begin{cases}
1 \quad 
\text{if } \ \forall d\in D(f,p), \  \gamma[f,p,d] = 1
\\
0 \quad\text{otherwise.}
\end{cases}
\end{equation}
For each family $f$ and each position $p$, create a binary variable $\beta[f, p]$ indicating whether family $f$ forms a blocking coalition with daycares $D(f, p)$. 
By the definition of stability, if family $f$ is matched to some assignment worse than the $p$-th assignment in $\succ_f$ (i.e., $\alpha[f,p] = 0$) and daycares $D(f, p)$ can accommodate $C(f)$ along with children with higher priority (i.e., $\gamma[f, p] = 1$), then family $f$ and daycares $D(f, p)$ form a blocking coalition. 
\begin{equation}
\label{var:beta}
\beta[f, p] = 
\begin{cases}
1 \quad \text{if } \alpha[f,p] = 0 \text{ AND } \gamma[f, p] = 1
\\
0 \quad\text{otherwise.}
\end{cases}
\end{equation}
We next explain how to encode desirable properties as constraints. 
Recall that for families with at least one child initially enrolled at a daycare $d\in D\setminus \{d_0\}$, we append initial enrollments $\omega(f)$ to the end of $\succ_f$. 
We can capture Individual Rationality with Constraint~\ref{cons:IR}: each family $f$ with initial enrollments (excluding dummy $d_0$) must be matched to an assignment in $\succ_f$. 
\begin{equation}
\label{cons:IR}
    \sum\nolimits^{|\succ_f|}_{p=1} x_{f, p} = 1 
\end{equation}
We enforce feasibility with the following two constraints.  The first constraint, labeled as Constraint~\ref{cons:fea:family}, specifies that each family can be matched with at most one assignment. 
The second constraint, labeled as Constraint~\ref{cons:fea:daycare}, mandates that for each daycare $d$ and each grade group $G_d \in G(d)$, the number of assigned children belonging to grade group $G_d$ must not exceed the transferable quota $Q(d, G_d)$. Here, 
$C(G_d) \subseteq C$ denotes the set of children belonging to grade group $G_d$.
\begin{equation}
\label{cons:fea:family}
    \sum\nolimits^{|\succ_f|}_{p=1} x_{f, p} \leq 1 \qquad \forall f\in F
\end{equation}
\begin{equation}
\label{cons:fea:daycare}
\sum\nolimits_{c\in C(G_d)} x_{c, d} \leq Q(d, G_d) \qquad \forall d\in D, \ \forall G_d \in G(d)
\end{equation}
\begin{theorem}
\label{theo:CP:stable}
Given an instance of daycare matching problem, create CP variables and constraints as defined above. 
There exists a stable outcome if and only if 
$\sum\nolimits_{f\in F} \sum\nolimits_{p \in [|\succ_f|]}\beta[f, p] = 0$ holds where $[k]$ means $\{1, 2, \cdots, k\}$.
\end{theorem}
We design two objective functions in our algorithm. The primary objective in Formula~\ref{cons:obj:bp} is to discover a feasible and individually rational matching with the least number of blocking coalitions. Let $\theta$ denote the minimum number of blocking coalitions among all such matchings. The secondary objective in Formula~\ref{cons:obj:max} is to determine a feasible and individually rational matching that incorporates the maximum number of matched children, subject to the condition that the number of blocking coalitions is no more than $\theta$. Note that we need to exclude the number of children who are matched to the dummy daycare $d_0$.
\begin{equation}
\label{cons:obj:bp}
    \min \sum\nolimits_{f\in F} \sum\nolimits_{p \in [|\succ_f|]}\beta[f, p] 
\end{equation}
\begin{equation}
\label{cons:obj:max}
     \max   \sum\nolimits_{c \in C} \sum\nolimits_{p=1}^{\mid\succ_{f,c}\mid} x_{c, p} - \sum\nolimits_{c \in C} \sum\nolimits_{p' \in P(c, d_0)} x_{c, p'}
\end{equation}

\begin{table*}[htb]
\centering
\resizebox{1.8\columnwidth}{!}{
\begin{tabular}{ccccccccccc}
    \hline
     & \multicolumn{2}{c}{Tama-21} & \multicolumn{2}{c}{Tama-22} & \multicolumn{2}{c}{Shibuya-21} & \multicolumn{2}{c}{Shibuya-22}
     & \multicolumn{2}{c}{Koriyama-22}
    \\
    & current & CP & current & CP & current & CP & current & CP & current & CP
    \\
    \hline
    \normalsize{\# matched children} & 558 & 560 & 464 & 470 & 1307 & 1307 & 1087 & 1087 & 979 & 1200
    \\
    \hline
    \normalsize{\# blocking coalition} & 26 & 0 & 2 & 0 & 0 & 0 & 0 & 0 & 1059 & 0
    \\
     \hline
\end{tabular}
}
\caption{Comparison with Status Quo Methods}
\label{table:outcome1}
\end{table*}

\begin{table*}[htb]
\centering
\resizebox{1.8\columnwidth}{!}{
\begin{tabular}{ccccccccccc}
    \hline
     & \multicolumn{2}{c}{Tama-21} & \multicolumn{2}{c}{Tama-22} & \multicolumn{2}{c}{Shibuya-21} & \multicolumn{2}{c}{Shibuya-22}
     & \multicolumn{2}{c}{Koriyama-22}
    \\
        & CP & CP-Ind & CP & CP-Ind & CP & CP-Ind & CP & CP-Ind & CP & CP-Ind
    \\
    \hline
    \normalsize{\# matched children} & 560 & 571 & 470 & 478 & 1307 & 1335 & 1087 & 1135 & 1200 & 1217
    \\
    \hline
    \normalsize{\# running time} & 9.6s & 12.5s & 1.7s & 3.0s & 18.1s & 1000s & 11.6s & 803s & 18.5s & 24.1s
    \\
     \hline
\end{tabular}
}
\caption{Impact of Indifferences. CP-Ind denotes the variant with indifferences allowed. CP-Ind found an optimal solution for all datasets within 1000s except for Shibuya-21.}
\label{table:indifference}
\end{table*}

\section{Empirical Evidence}
In this section, we evaluate the performance of our new algorithm through experiments on several real-life data sets. Our experiment comprises two parts:  
Firstly, we compare our algorithm with currently deployed algorithms in terms of two principal objectives of the daycare matching problem: the number of matched children and the number of blocking coalitions.
Experimental results show that our algorithm consistently outperforms the existing methods, especially a commercial software that dominates the current daycare matching market in Japan.
Secondly, we examine the effect of different factors on the number of matched children, 
such as allowing indifferences in daycares' priorities and permitting a small number of blocking coalitions. 

\textbf{Methodology} We implemented our CP model using the Google OR-Tools API. All the experiments were performed on a laptop equipped with an M1-max CPU and 32GB of RAM. We wrote our code in Python and adapted it for the CP-SAT solver offered in the OR-Tools package, which is a powerful and award-winning solver for constraint programming problems.
Please refer to the international constraint programming competition (MiniZinc Challenge) for more details.

The data sets we tested are provided by three municipalities: Shibuya, a major commercial and finance center in Tokyo; Tama, a suburban city located in the west of Tokyo; and Koriyama, a large city located in the northern region of Japan.  
While our CP algorithm is capable of handling transferable quotas, it is important to note that we only have access to a relatively small data set that allows for transferable quotas. Unless explicitly stated otherwise, the following experimental results will focus on data sets that include only two features: siblings and initial enrollments. The number of children participating in these data sets varies from 550 to 1589 and the number of daycares varies from 33 to 86. We provide detailed descriptions of these data sets in Appendix.

\subsection{Experiment 1: Comparison with Status Quo Methods} 
For Shibuya and Koriyama data sets, the status quo algorithm is provided by a commercial and non-disclose software. However, Tama also acquired the same software but ultimately disregarded it due to its unsatisfactory performance, and currently employs a manual approach instead.
%
In Table~\ref{table:outcome1},
we summarize the results of experiments comparing these methods. Notably, the CP algorithm not only increases the number of matched children by up to $23\%$, but also generates outcomes with the minimum number of blocking coalitions, ensuring a more  satisfactory outcome with respect to fairness and non-wastefulness. In the next experiment, we further show that it is still possible to decrease the number of unmatched children while maintaining stability.

\subsection{Experimental 2: Effect of Different Factors on the Outcomes} 
Next, we present a summary of the findings regarding the impact of different factors on the number of matched children.
1) \textbf{Indifferences}
In the current daycare matching process, each municipality employs a distinct and intricate scoring system to establish the priority order for children. However, tie-breakers are frequently employed by introducing a small fractional value to derive a strict priority ordering. An alternative approach we propose is to remove the small fractional values and allow for indifferences in daycare priorities. 
Our research summarized in Table~\ref{table:indifference} reveals that incorporating indifferences can lead to a substantial decrease in the number of unmatched children, 
while still maintaining stability.
It is important to note that computing such a stable matching with maximum matches may require considerably more time in certain cases.
2) \textbf{Blocking Coalition} 
Throughout our tests on all the data sets, we consistently achieved stable outcomes, wherein no blocking coalitions were present. However, we wanted to explore the possibility of allowing a small number of blocking coalitions to potentially benefit more children.
Surprisingly, our investigations revealed that incorporating a small number of blocking coalitions, such as 5, did not necessarily lead to a significant increase in the number of matched children. In contrast, a larger number of blocking coalitions, such as 100, showed a noticeable impact on increasing the number of matched children. 
However, it is important to note that as the number of permitted blocking coalitions increases, the computational time required to find a feasible solution also increases significantly.
3) \textbf{Transferable Quotas} 
We only have one data set in which transferable quotas are allowed. Although the use of transferable quotas could lead to an increase of matched children in theory, 
we cannot draw a clear conclusion on the effect of transferable quotas due to the small size of the data set.

\section{Conclusion}
In this paper, we describe three key features of the daycare matching problem in Japan, which consolidate and integrate other important matching models such as hospital-doctor matching with couples. We propose a practical algorithm based on constraint programming (CP) that maximizes the number of children matched to daycare centers while ensuring stability, a desirable property in the context of matching problems under preferences.
Through experimentation, we have discovered that our CP algorithm holds great potential in significantly improving the process of matching children to daycare centers when compared to the current methods employed in practice. This algorithm presents a promising approach to addressing the daycare matching problem and has the capacity to shed light on similar matching problems.
%

\section{Acknowledgments}
This work was partially supported by JST ERATO Grant Number JPMJER2301 and JSPS KAKENHI Grant Number JP21H04979, Japan.

\bibliography{aaai24}

\begin{thebibliography}{25}
\providecommand{\natexlab}[1]{#1}

\bibitem[{Abdulkadiro{\u{g}}lu and S{\"o}nmez(2003)}]{AbSo03b}
Abdulkadiro{\u{g}}lu, A.; and S{\"o}nmez, T. 2003.
\newblock School Choice: A Mechanism Design Approach.
\newblock \emph{American Economic Review}, 93(3): 729--747.

\bibitem[{Ashlagi, Braverman, and Hassidim(2014)}]{ABH14a}
Ashlagi, I.; Braverman, M.; and Hassidim, A. 2014.
\newblock Stability in Large Matching Markets with Complementarities.
\newblock \emph{Operations Research}, 62(4): 713--732.

\bibitem[{Bir{\'o}, Manlove, and McBride(2014)}]{BMM14a}
Bir{\'o}, P.; Manlove, D.~F.; and McBride, I. 2014.
\newblock The hospitals/residents problem with couples: Complexity and integer
  programming models.
\newblock In \emph{International Symposium on Experimental Algorithms}, 10--21.
  Springer.

\bibitem[{Correa et~al.(2022)Correa, Epstein, Epstein, Escobar, Rios, Aramayo,
  Bahamondes, Bonet, Castillo, Cristi, Epstein, and Subiabre}]{CEE+22a}
Correa, J.; Epstein, N.; Epstein, R.; Escobar, J.; Rios, I.; Aramayo, N.;
  Bahamondes, B.; Bonet, C.; Castillo, M.; Cristi, A.; Epstein, B.; and
  Subiabre, F. 2022.
\newblock School choice in Chile.
\newblock \emph{Operations Research}, 70(2): 1066--1087.

\bibitem[{Dur, Morrill, and Phan(2022)}]{DMP22a}
Dur, U.; Morrill, T.; and Phan, W. 2022.
\newblock Family Ties: School Assignment with Siblings.
\newblock \emph{Theoretical Economics}, 17(1): 89--120.

\bibitem[{Gale and Shapley(1962)}]{GaSh62a}
Gale, D.; and Shapley, L.~S. 1962.
\newblock College admissions and the stability of marriage.
\newblock \emph{The American Mathematical Monthly}, 69(1): 9--15.

\bibitem[{Garey and Johnson(1979)}]{GaJo79a}
Garey, M.~R.; and Johnson, D.~S. 1979.
\newblock Computers and Intractability: A Guide to the Theory of
  {NP}-Completeness.

\bibitem[{Geitle et~al.(2021)Geitle, Johnsen, Ruud, Fagerholt, and
  Julsvoll}]{GJR+21a}
Geitle, A.~H.; Johnsen, {\O}.~K.; Ruud, H. F.~E.; Fagerholt, K.; and Julsvoll,
  C.~A. 2021.
\newblock Kindergarten allocation in Norway: An integer programming approach.
\newblock \emph{Journal of the Operational Research Society}, 72(7):
  1664--1673.

\bibitem[{Hamada et~al.(2017)Hamada, Hsu, Kurata, Suzuki, Ueda, and
  Yokoo}]{HHKS+17a}
Hamada, N.; Hsu, C.; Kurata, R.; Suzuki, T.; Ueda, S.; and Yokoo, M. 2017.
\newblock Strategy-proof school choice mechanisms with minimum quotas and
  initial endowments.
\newblock \emph{Artificial Intelligence}, 249: 47--71.

\bibitem[{Hatfield and Milgrom(2005)}]{HaMi05a}
Hatfield, J.~W.; and Milgrom, P.~R. 2005.
\newblock Matching with contracts.
\newblock \emph{American Economic Review}, 95(4): 913--935.

\bibitem[{Kamada and Kojima(2023)}]{KaKo23a}
Kamada, Y.; and Kojima, F. 2023.
\newblock Fair Matching under Constraints: Theory and Applications.
\newblock \emph{The Review of Economic Studies}.

\bibitem[{Kojima, Pathak, and Roth(2013)}]{KPR13a}
Kojima, F.; Pathak, P.~A.; and Roth, A.~E. 2013.
\newblock Matching with couples: Stability and incentives in large markets.
\newblock \emph{The Quarterly Journal of Economics}, 128(4): 1585--1632.

\bibitem[{Manlove, McBride, and Trimble(2017)}]{MMT17a}
Manlove, D.~F.; McBride, I.; and Trimble, J. 2017.
\newblock ``{A}lmost-stable'' matchings in the Hospitals/Residents problem with
  Couples.
\newblock \emph{Constraints}, 22(1): 50--72.

\bibitem[{McDermid and Manlove(2010)}]{McMa10a}
McDermid, E.~J.; and Manlove, D.~F. 2010.
\newblock Keeping partners together: algorithmic results for the
  hospitals/residents problem with couples.
\newblock \emph{Journal of Combinatorial Optimization}, 19(3): 279--303.

\bibitem[{Nguyen and Vohra(2018)}]{NgVo18a}
Nguyen, T.; and Vohra, R. 2018.
\newblock Near-feasible stable matchings with couples.
\newblock \emph{American Economic Review}, 108(11): 3154--69.

\bibitem[{Okumura(2019)}]{Okum19a}
Okumura, Y. 2019.
\newblock School choice with general constraints: a market design approach for
  the nursery school waiting list problem in {J}apan.
\newblock \emph{The Japanese Economic Review}, 70(4): 497--516.

\bibitem[{Reischmann, Klein, and Giegerich(2021)}]{RKG21a}
Reischmann, T.; Klein, T.; and Giegerich, S. 2021.
\newblock An iterative deferred acceptance mechanism for decentralized, fast
  and fair childcare assignment.
\newblock \emph{ZEW-Centre for European Economic Research Discussion Paper},
  (21-095).

\bibitem[{Ronn(1990)}]{Ronn90a}
Ronn, E. 1990.
\newblock {NP}-complete stable matching problems.
\newblock \emph{Journal of Algorithms}, 11(2): 285--304.

\bibitem[{Roth(1984)}]{Roth84a}
Roth, A.~E. 1984.
\newblock The evolution of the labor market for medical interns and residents:
  a case study in game theory.
\newblock \emph{Journal of Political Economy}, 92(6): 991--1016.

\bibitem[{Roth(2008)}]{Roth08a}
Roth, A.~E. 2008.
\newblock Deferred acceptance algorithms: History, theory, practice, and open
  questions.
\newblock \emph{International Journal of Game Theory}, 36(3): 537--569.

\bibitem[{Roth and Sotomayor(1992)}]{RoSo92a}
Roth, A.~E.; and Sotomayor, M. 1992.
\newblock Two-sided matching.
\newblock \emph{Handbook of game theory with economic applications}, 1:
  485--541.

\bibitem[{Sun et~al.(2023)Sun, Takenami, Moriwaki, Tomita, and Yokoo}]{STM+23a}
Sun, Z.; Takenami, Y.; Moriwaki, D.; Tomita, Y.; and Yokoo, M. 2023.
\newblock Daycare Matching in {J}apan: Transfers and Siblings.
\newblock In \emph{Thirty-Seventh {AAAI} Conference on Artificial Intelligence,
  {AAAI}}, 14487--14495.

\bibitem[{Suzuki et~al.(2023)Suzuki, Tamura, Yahiro, Yokoo, and
  Zhang}]{STY+23a}
Suzuki, T.; Tamura, A.; Yahiro, K.; Yokoo, M.; and Zhang, Y. 2023.
\newblock Strategyproof Allocation Mechanisms with Endowments and {M}-convex
  Distributional Constraints.
\newblock \emph{Artificial Intelligence}, 315: 103825.

\bibitem[{Suzuki, Tamura, and Yokoo(2018)}]{STY18a}
Suzuki, T.; Tamura, A.; and Yokoo, M. 2018.
\newblock Efficient allocation mechanism with endowments and distributional
  constraints.
\newblock In \emph{Proceedings of the 17th International Conference on
  Autonomous Agents and MultiAgent Systems}, 50--58.

\bibitem[{Veski et~al.(2017)Veski, Bir{\'o}, P{\"o}der, and Lauri}]{VBPL17a}
Veski, A.; Bir{\'o}, P.; P{\"o}der, K.; and Lauri, T. 2017.
\newblock Efficiency and fair access in kindergarten allocation policy design.
\newblock \emph{The Journal of Mechanism and Institution Design}, 2(1):
  57--104.

\end{thebibliography}

\end{document}